# A new path algorithm for the weighted multi-graphs WMGPA: application to the Direct Topological Method

Abderrahmane Euldji[1], Abderrahim Tienti[2] and Amine Boudghene Stambouli[3]

[1] **National Institute of Telecommunications and ICT ( INT-TIC ), Oran, Algeria**
*euldji2005@hotmail.com*

[2] **National Institute of Telecommunications and ICT ( INT-TIC ), Oran, Algeria**
*atienti@yahoo.fr*

[3] **Department of Electricity, University of Science and Technology of Oran USTO, Oran, Algeria**
*aboudghenes@gmail.com*

**Abstract**
The aim of this paper is to present an algorithm which gives all the possible paths that start from a specific node to another of a weighted multi-graph. This algorithm is intended to be applied for the direct topological method.

## 1. Introduction

The path notion is one of the most important graph traversal concepts. To find a path in a graph, two vertices shall be specified as a beginning and an ending vertex. The path is the chain that starts from the beginning vertex until it reaches the ending node, such that every vertex in this chain shall not be crossed twice. As a result, a path is a simple graph that does not contain any isolated node.

The path can be used to describe certain characteristics of a graph, or to solve other problems related to this latter. It can be used for decisional purposes, or for optimization needs. In general, many topics in the graph theory are based upon the path notion.

For this purpose, many efforts have been done in order to give an "optimized" algorithmic form that finds all the possible paths that starts from a specific node to another in a graph; the most well-known all paths extractor algorithms are the BFS (Breadth First Search), and the DFS (Depth First Search) and their derivative algorithms [1][2][3][4][5]. But since there are multiple classifications of graphs, then a path algorithm made for a specified class of graphs may or may not be used for the other classes of graph.

For electrical circuits -in which this paper is established for- the DFS at a first sight may appear as the appropriate procedure to find paths of the electrical circuit' associated graph, but since they are considered as weighted multi-graphs, then DFS becomes ineffective (incompetent). Hence, this paper proposes a path algorithm suited for the electrical circuits' analysis, and that can be used specifically for the direct topological method DTM; it is called: WMGPA.

Before everything, it is better suited to talk about the selected electrical circuits' analysis method in this paper, which is the direct topological method DTM.

## 2. Direct Topological Method DTM

The methods of analysis of the electrical circuits are divided into two main classes:

- The algebraic methods: based on the resolution of the circuits equations, either by conventional computations, or through the intermediary of the signal flow graph.

- The topological methods: they study the circuit's structure (topology) to deduct the expression of the circuit's functions.

The direct topological method (D.T.M) -as its name indicates- is a method of analysis of the permanent linear circuit networks, which permits to write directly the terms of the looked for circuit functions, under their definitive compact form, by visual examination of the studied circuit topology. While reducing the steps to get the result without complex computations, by simple application of the topological rules. Hence, it permits to solve a problem of circuit's analysis with minimum effort and economy of time in comparison with the other methods.

D.T.M is based on the graph theory definitions, since it is among the topological methods.

Any circuit function is rational; D.T.M intends to write separately the numerator and the denominator, while with the other analysis methods, only the circuit function is well determined. The denominator is called the topological determinant D; for the case of transfer functions, the numerator is called the topological transfer numerator N, both of these two parameters are called: the circuit's topological functions [6].

This method analyzes many types of electrical networks: passive or active circuits, circuits with dependent or independent sources, circuits containing one or many mutual inductances as well as ideal transformers. This method analyzes also circuits with distributed parameters as transmission lines. All of these circuits require -in one or many phases of their analysis process- the analysis of RLC circuits.

Before introducing how to analyze RLC circuits by the D.T.M, one should know some graph theory definitions.

## 3. Graph theory

Graph theory is the study of the properties of graph structures. It provides us with a language with which to talk about graphs. The key to solve many problems is identifying the fundamental graph-theoretic notion underlying the situation and then using classical algorithms to solve the resulting problem.

Graph theory is very wide domain, in constant evolution in either fundamental or applicative research; the applications are very numerous, which justifies an important research in algorithmic.

The graph theory offers a certain educational interest on the other hand. Indeed, the definitions are simple, and real problems of research can be posed as "mathematical games" whose playful formulation can cover big difficulties.

As the graph models very numerous situations, the proposed problems are of more "natural" form [7].

The multiplicity of the applications also explains the variety of the definitions, or of the variants of definitions. So an article of graph theory always must "fix the definitions"[8].

3.1 Relative definitions to topological graphs

*1) node (vertex):* A node "v" is an extremity point or an intersection point between branches.

**2) Branch:** A branch "e" is a link between two nodes; there are two types of branches, oriented branches (it is called in this situation by arc) and non-oriented branches (it is called in this situation by edge) .

**3) Graph:** A graph "G" (v, e) is constituted from "v", a nonempty set of nodes, and from "e", a set of branches, where each branch from "e" is a connection between two nodes from "v"; nodes number is denoted $|v|$, and the branches number is $|e|$ [3].

There exist many classifications for graphs; such as:

**a) Embedded vs. Topological graph:** A graph is embedded if the vertices and branches have been assigned geometric positions. Thus any drawing of a graph is an embedding, which may or may not have algorithmic significance. Occasionally, the structure of a graph is completely defined by the geometry of its embedding.

**b) Simple graph vs. Multi-graph:** A simple graph is a graph having no loops or multiple branches. In this case, each branch in E(G) can be specified by its endpoints u; v in V(G).

In contrast, a multi-graph is a graph that it may contain loops and at least one multiple branches (*two adjacent vertices are connected via multiple branches*)

**c) Directed graph (digraph / oriented graph):** a directed graph (digraph) is a graph where at least one branch in it is oriented (arcs).

**d) Weighted graph:** In weighted graphs, each branch (or vertex) of G is assigned a numerical (or symbolical) value, or weight.

**e) Labeled vs. Unlabeled graph:** In labeled graphs, each vertex is assigned a unique name or identifier to distinguish it from all other vertices [7].

**4) Node's degree of a non-oriented graph:** it is the number of non-looped branches connected to this node.

3.2 Links and attributes

**1) Loop:** There is a loop when the branch extremities correspond to the same node.
**2) Cycle:** A cycle is a closed path. A loop is a cycle with a single node and branch.

**3) Supplement:** supplement of a branch set [$e_1$, $e_2$,…, $e_n$] in a graph "G" is the resulted graph from the elimination of

these branches, each one is followed by the coincidence of correspondent extremities.

**4) Connected graph:** A graph "G" is called connected, if it is possible to find at least a path joining two arbitrary nodes from it. A non-connected graph is a degenerated one [6].

**5) Transfer Graph:**

**a) Transfer cycle:** There are two distinct branches from a graph; "g" is called: the transmitter, and "h" is called the receptor; these two branches form a couple K = (g , h) called the transfer couple. Each cycle from a graph containing at the same time "g" and "h" is a transfer cycle joining "g" and "h", noted (g → h). A graph containing transfer couples is a transfer graph.

**b) Sign of the cycle:** The sign of the transfer cycle joining g to h is +1 or -1 depending on whether g and h are in the same direction or in an opposite direction in the cycle.

## 4. Analysis of RLC circuits by the D.T.M

The analysis process is subdivided into many steps which depend on the specified network function: if it is a transfer function, than a calculus of a topological transfer numerator and a topological determinant is required. If it is a driving point function (input function) than two topological determinants are needed.

Both of the circuit's topological functions are calculated by following a set of steps, these steps vary according to the complexity of their associated graphs, if the graph is complex, then further rules and theorems are executed.

These rules and theorems are specified to extract either the topological determinant or the topological transfer numerator, for either of these two situations, some of these rules and theorems are elementary (like rule n°1 and theorem n°1), some are executed for special topologies, others are general.

In order to identify the procedure to obtain the topological functions, the class of these graphs shall be identified as well. The classification is important. The class of the graph implies the procedure to process it (this includes the algorithm(s) to apply). The two previously mentioned graphs to be extracted are both labeled and weighted topological multi-graphs. In the case of the transfer numerator's graph, it is considered as a directed graph, in contrast with the topological determinant's graph (it is a non-directed graph).

### 4.1 Topological Determinant D

To calculate the topological determinant of a RLC circuit, we consider the graph of its non-excited circuit (hence it is non-directed graph). The excitation sources in the circuit have to be replaced by short-circuits for voltage generators, and by opened circuits for current generators [6].

Afterwards, the circuit's topology is processed; it includes also the verification of the circuit's complexity.

If the graph is simple, than a table of elementary topological determinants (TED) is consulted, if this table does not contain this graph, or the graph is complex, than further rules and theorems are needed.

In this paper, we introduce only two rules [9]:

**1) Rule n° 1:** An inductance "L" must be considered as impedance "LS', and a capacitance "C" must be considered as admittance "CS". A resistance could be considered as impedance "R", or admittance "G".

**2) Rule n° 3:** The determinant of a degenerated circuit is null.

A circuit is degenerated when it is not connected, or when it contains a cycle which all its branches have a null impedance. This cycle corresponds to a degenerated node of the graph of the circuit.

### 4.2 Transfer Numerator

To get the transfer function, we must have besides the determinant the transfer numerator. Contrary to the determinant that only depends on the non-excited circuit, the transfer numerator depends not only on the placement of the excitation source, but also to the one of the output response. Therefore, we have then to consider the circuit's graph with the excitation source and the output response. This graph differs from the non-excited circuit's graph by the existence of a transmitter branch, which represents the excitation source, and a receiver branch, which represents the output response, which both of them are oriented branches. Therefore, this is the transfer graph of a single transfer couple; it can contain several transfer cycles joining the source excitation to the output response.
In this paper, we introduce only one theorem.

**Theorem n°3:** The transfer numerator of a RLC circuit is the algebraic sum of the circuit's transfer cycles values [9].

The value of the transfer cycle, as a definition, equals the multiplication of:

- The cycle's sign.
- Admittances being part of the cycle.
- The cycle's supplement determinant of the circuit.

## 5. Specification of the data structures

As already said, the graphs classification have a major impact on the choice of the procedure/algorithm to be applied. This also have an important role for the choice of the most appropriate data structures to be selected.

Since the previously mentioned graphs to be extracted are both labeled and weighted topological multi-graphs, then the proposed data structures in this case are as follows:

**a) Vertex list:** This list holds the data of every vertex in the graph; they are the vertex's label (numerical value) and degree.

**b) Non-looped branches list:** This list holds the data of all the non-looped branches in the graph.

**c) Looped branches list:** Since the multi-graphs may contain looped branches, thus it will be much more appropriate to put them in their own list.

For both non-looped and looped branches lists, the vertex's labels in which these branches are connected to shall be included in both of them.

The branch's orientation defines whether the branch is directed (true) or non-directed (false). It is to note that for the directed branches, the order of the branch's vertices always indicates the positive direction.

The name of the branch is composed from the nature of the component in which it is represented by this branch, plus the index of the component which differs between the components with the same nature. For example: if the branch represents a resistor with the index 1, then its name will be: R1.

According to rule n°1, the RLC components have an affinity to be impedances or admittances, so this affinity will be indicated as the category of the branch. Hence, every branch shall have a category name that is obtained from the association of the component's affinity, and an index to differ between components with the same category. For example: if G have the following components: R1, L1, C1 and C2, their category names are (respectively): Z1, Z2, Y1 and Y2.

The weight of the branch is the category of the component. It can be defined by both of the branch's category name and type name.

The previous mentioned data structures for the branches are made to store the data of each branch without to consider its connection with the rest of the branches, this connection gives a look at the topology of the graph. So to illustrate it, the adjacency-incidence list is defined.

**d) Adjacency-incidence list:** In this representation, the graph's data are represented by using linked lists to store the entire incident branches (looped branches are excluded) to each vertex in the graph. In the same process, the neighbored vertices to each actual vertex which is connected to those branches are stored as well. Typically to construct this list, all the branches of the graph -through the non-looped branches list- are swept (in coordination with the vertex list), and the adjacency-incidence list is updated. As a result, one can identify the adjacent branches to a specific branch, and in the same time one can identify all the incident branches to a specific node.

**Example**: Let us make the adjacency-incidence list of the following circuit (Fig.1).

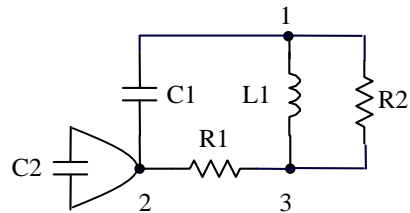

Fig. 1 Example of a circuit.

The graph associated to this circuit is called G.

This graph has one looped branch, so the looped branches list is as presented in Table 1.

Its Non-looped branches list as presented in Table 2.

Table 1: Looped branches list of graph G

| Vertex Label | Component's Category Name | Component's Nature Name |
|---|---|---|
| 2 | Y2 | C2 |

Table 2: Non-looped branches list of graph G

| Branch's 1st Node | Branch's 2nd Node | Branch's Orientation | Branch's Nature Name | Branch's Category Name |
|---|---|---|---|---|
| 1 | 2 | false | C1 | Y1 |
| 1 | 3 | false | L1 | Z3 |
| 2 | 3 | false | R1 | Z1 |
| 1 | 3 | false | R2 | Z2 |

This graph has three nodes, so the vertex list is as illustrated in Table 3.

Table 3: Vertex list of graph G

| Vertex Label | Vertex Degree |
|---|---|
| 1 | 3 |
| 2 | 2 |
| 3 | 3 |

Then, the Adjacency-incidence list is as presented in Table 4.

Table 4: Adjacency-incidence list of graph G

| Vertex data | | Adjacency-incidence list data |
|---|---|---|
| Label | Degree | Adjacent Vertex → Incident Branch's Nature Name / Category Name |
| 1 | 3 | 2 → C1/Y1 ; 3 → L1/Z3 ; 3 → R2/Z2 |
| 2 | 2 | 1 → C1/Y1 ; 3 → R1/Z1 |
| 3 | 3 | 1 → L1/Z3 ; 1 → R2/Z2 ; 2 → R1/Z1 |

Now, let us discuss the path's algorithm.

## 6. The weighted multi-graph's path algorithm WMGPA

The weighted multi-graph path algorithm WMGPA needs as a principal element for its development the adjacency-incidence list, because this list is simply a list of very elementary paths which contain only the beginning and ending nodes. From this remark, the WMGPA is as follows:

```
WMGPA ( BgnNd , EndNd, GRAPH )
{
    if ( BgnNd == EndNd ) {
        Printf ( "Error ! Beginning node shall be different than the ending node.") ;     }
    else {
    TempPathCounter = Degree of BgnNd ;
    Creating path copies containg BgnNd in the temporary path array TempPath with a number equal to TempPathCounter ;

    for ( i = 0 ; i < TempPathCounter ; i++ )
    {
        Search in the Adjacency-incidence List of GRAPH for a match between data of the last node in TempPath[i];

        if ( last node in TempPath[i] == EndNd) {
            Save TempPath[i] in the PathList array ;
            Delete TempPath[i] ;
            Decrement TempPathCounter by 1 ;     }
        else if ( last node degree in TempPath[i] == 1 ) {
            Delete TempPath[i] ;
            Decrement TempPathCounter by 1 ;     }
    }

    for ( i = 0 ; i < TempPathCounter ; i++ )
    {
        Counter = Degree of the last node in TempPath [i];

        Create path copies from TempPath[i] , with a number equal to Counter and store them in TempPathAnalyzer ;

        Search in the Adjacency-incidence List for a match between data of the last node in TempPathAnalyzer ;

        Save the data of each neighboured element to the last node in the last position of each path stored in TempPathAnalyzer ;

        for ( j = 0 ; j < Counter ; j++ )
        {
            if ( last node in TempPathAnalyzer[j] is EndNd )
            {
                Save TempPathAnalyzer[j] in the PathList;
                Mark TempPathAnalyzer[j] as unwanted ;     }
            else if ( last node in TempPathAnalyzer[j] is repeated in it ) {
                Mark TempPathAnalyzer[j] as unwanted ;     }
            else if ( Degree of the last node in TempPathAnalyzer[j] == 1 ) {
                Mark TempPathAnalyzer[j] as unwanted ;     }
            else {
                Save TempPathAnalyzer [j] in TempPath ;
                Increment TempPathCounter by 1 ;
                Mark TempPathAnalyzer [j] as unwanted ;     }
        }
    }
  }
}
```

6.1 Description of the algorithm

This algorithm is intended to extract all the possible paths which start with the beginning vertex BgnNd and the ending vertex EndNd from a graph GRAPH.

As a first step, the algorithm processes the before mentioned vertices to make sure that they are not the same. Afterwards, it builds an array that stores the temporary paths called TempPath, and an array that stores the positively verified paths PathList.

There are two stages in this process:

As a first stage, TempPath stores BgnNd, then it starts to process -through the "adjacency-incidence" list- the neighboured nodes to this latter (without to forget their connecting branch). If the node to be processed is effectively EndNd, then this temporary path is verified as a complete path and it is stored in PathList (and eliminated from the temporary path array). If this is not the case, then if this node has a degree of one, then this temporary path is rejected due to the fact that this node can not pass to any other node anymore, then it can not pass to the ending node. If this condition is not correct, then this path is verified as temporary, and it may lead in the next "hop" to the ending node.

In the second stage, the last vertex is selected in order to get its neighboured vertices (and the branch connecting these two) by the mean of the adjacency-incidence list, then, another array arises which is TempPathAnalyzer, which presents a table that stores the established temporary "temporary paths" from the TempPath array, and the same procedure as in the first stage is set with an extra condition, which is the test of the pre-existence of this additional node in the temporary path, which means in other words that this node is repeated, and so it results in a rejection of this possibility. This process keeps looping until the temporary path array becomes void, at this state, all the possibilities have been considered and the algorithm finishes.

## 7. Application of the weighted multi-graph path algorithm in the DTM

The direct topological method may use WMGPA in many of its stages. One of its applications which is explicit and presents one of its most useful benefits is its ability to study the connectivity of the graph. As already mentioned in part III.B.4, if there is no way to find any path joining two nodes in a graph, then this graph is degenerated and as rule n°3 states, the determinant of a degenerated graph is null.

One of its implicit uses is when it is molded to become the cycle's algorithm, this algorithm can be used –in one of its applications– as a transfer cycle to calculate the topological transfer numerator according to the third theorem of this method (part IV.B).

## 8. Conclusion

In this paper, we proposed a new path algorithm that finds all the possible paths that start from a specific node to another in a graph. This algorithm is suited to the weighted multi-graphs; it can be considered as an extended DFS algorithm. It can be used to develop algorithm for DTM in order to analyze electrical circuits.

**Abderrahmane Euldji** is a state engineer from the National Institute of Telecommunications and ICT (INT-TIC) since 2008. He is preparing for the magister degree in telecommunications, option: network systems and ICT. His current interest is computer science and algorithmic.

**Abderrahim Tienti** had his state engineer degree in telecommunications in 1983 from the National Institute of Telecommunications and ICT, a DEA in automatics from ENSIEG of INPG (Gronoble, France) in 1990 and his magister degree in robotics from University of Sciences and Technology of Oran (Algeria) in 1998. He is a lecturer at INT-TIC. His current interest is the analysis and synthesis of electrical circuits (especially in DTM), electronic switching systems, signaling system SS7, xDSL, ISDN. He has contributed in many papers in many national and international conferences.

**Amine Boudghene Stambouli** is a graduate of the University of Sciences and Technology of Oran (Algeria) in 1983. He received his master's degree in modern electronics (1985) and his PhD in optoelectronics (1989) at the University of Nottingham in England. He joined the University of Sciences and Technology of Oran in 1989. His studies started in the field of High Field Electroluminescence and optoelectronics and lately changed to environmental friendly production of energy. His research interests include at present: Photovoltaics, Fuel cells, hybrid systems, and environment impacts. He is a full Professor of optoelectronics and material science for environment and energy applications at the Department of Electronics. Prof. Amine Boudghene Stambouli is United Nations consultant (Index 382958), member of many scientific and industrial organizations and director of several doctoral courses. He served as the head of the electronics department, vice rector of the university for almost two years and later the president of the scientific council of the electrical and electronics engineering faculty. Prof. Amine Boudghene Stambouli has been chairing five international conferences in the field of electrical engineering and was chairman at numerous sessions of international conferences. His studies are documented by 1 book, 3 polycopies and several papers mainly published on International

Journals and on Proceedings of International and National Conferences. He is a reviewer and an Editorial Board Membership of several International Journals. He is actively collaborating with research group world wide (Algeria, Italy, Japan, France, USA, Turkey, England, Saudi Arabia, Jordan and Syria). He was Co-responsible, with Pr. Enrico travera, of the research team « Photovoltaics and fuel cells » between the university of Roma Tor Vergata and the university of Sciences and Technology of Oran. He was awarded the prise of the best publication of the year 2009, delivered by CDER and the ministry of higher education and scientific research of Algeria. He is Co-responsible (Algerian side) of the Sahara Solar Breeder (SSB) project along with Pr. Koinuma (Japan side). Founder of the Sahara Solar Breeder Foundation.